\documentclass[sigconf]{acmart}
\usepackage{booktabs} 

\usepackage{subfigure}
\usepackage{amsfonts} 
\usepackage{amsmath}  
\usepackage{multirow}

\setcopyright{rightsretained}

\acmDOI{}

\acmISBN{}

\acmConference[]{}{}{}
\acmYear{1997}
\copyrightyear{2018}

\acmArticle{4}
\acmPrice{15.00}

\begin{document}
\title{Eye movement simulation and detector creation to reduce laborious parameter adjustments}
\titlenote{Produces the permission block, and
  copyright information}
\subtitle{}
\subtitlenote{}

\author{Wolfgang Fuhl}
\affiliation{%
\institution{University Tuebingen, Perception Engineering}
	\streetaddress{Sand 14}
	\city{Tuebingen}
	\state{Baden-Wuerttemberg}
	\postcode{72076}
}
\email{wolfgang.fuhl@uni-tuebingen.de}

\author{Thiago Santini}
\affiliation{%
	\institution{University Tuebingen, Perception Engineering}
	\streetaddress{Sand 14}
	\city{Tuebingen}
	\state{Baden-Wuerttemberg}
	\postcode{72076}
}
\email{thiago.santini@uni-tuebingen.de}

\author{Thomas Kuebler}
\affiliation{%
	\institution{University Tuebingen, Perception Engineering}
	\streetaddress{Sand 14}
	\city{Tuebingen}
	\state{Baden-Wuerttemberg}
	\postcode{72076}
}
\email{thomas.kuebler@uni-tuebingen.de}

\author{Nora Castner}
\affiliation{%
	\institution{University Tuebingen, Perception Engineering}
	\streetaddress{Sand 14}
	\city{Tuebingen}
	\state{Baden-Wuerttemberg}
	\postcode{72076}
}
\email{castnern@informatik.uni-tuebingen.de}

\author{Wolfgang Rosenstiel}
\affiliation{%
	\institution{University Tuebingen, Technical Computer Science}
	\streetaddress{Sand 14}
	\city{Tuebingen}
	\state{Baden-Wuerttemberg}
	\postcode{72076}
}
\email{Wolfgang.Rosenstiel@uni-tuebingen.de}

\author{Enkelejda Kasneci}
\affiliation{%
	\institution{University Tuebingen, Perception Engineering}
	\streetaddress{Sand 14}
	\city{Tuebingen}
	\state{Baden-Wuerttemberg}
	\postcode{72076}
}
\email{Enkelejda.Kasneci@uni-tuebingen.de}

\renewcommand{\shortauthors}{}

\begin{abstract}
Eye movements hold information about human perception, intention and cognitive state. Various algorithms have been proposed to identify and distinguish eye movements, particularly fixations, saccades, and smooth pursuits. A major drawback of existing algorithms is that they rely on accurate and constant sampling rates, impeding straightforward adaptation to new movements such as micro saccades. We propose a novel eye movement simulator that i) probabilistically simulates saccade movements as gamma distributions considering different peak velocities and ii) models smooth pursuit onsets with the sigmoid function. This simulator is combined with a machine learning approach to create detectors for general and specific velocity profiles. Additionally, our approach is capable of using any sampling rate, even with fluctuations. The machine learning approach consists of different binary patterns combined using conditional distributions. The simulation is evaluated against publicly available real data using a squared error, and the detectors are evaluated against state-of-the-art algorithms.
\end{abstract}

%
%



\maketitle

\section{Introduction}
Eye movements hold valuable information about a subject, her intension and cognitive states~\cite{braunagel2017online,kubler2017subsmatch} and are also important for the diagnosis of defects and diseases of the eyes (many examples can be found in \cite{leigh2015neurology}). Therefore, the detection and differentiation of eye movement types has to be accurate. Most algorithms for eye movement detection apply different dispersion, velocity or acceleration thresholds and validate the detected eye movements based on their duration. This approach seems to be unsatisfactory~\cite{andersson2017one} at its current state. This is partially due to instable or even dynamic sampling rates of eye tracking devices, task specific sources of noise, the interpolation method applied to the data by the eye tracker, and several more~\cite{cornelissen2002eyelink,duchowski2002breadth}. Depending on the task at hand, different thresholds are proposed in the literature~\cite{holmqvist2011eye}. It is especially difficult to adjust these thresholds for inconsistent sampling rates and noise which is not annotated by the eye tracker. Some commercial eye-tracker differ between tracking the eye and pupil and re-detecting them after a tracking loss, where the latter requires significantly more processing time and thus results in a decreased frame rate. Therefore, the identification of eye movements is still a difficult task; it complicates to confidently generalize research findings across experiments~\cite{andersson2017one}. 

Classifying eye movements is the process of separating different intervals in the gaze data to certain oculomotor and cognitive processes. For example visual perception during a saccade is severely limited~\cite{rayner1998eye,kliegl1981reduction}. In constast to these very fast movements, perception is working during the (much slower) pursuit of a moving object~\cite{rashbass1961relationship}. Another important eye movement is blinking. While it is not primarily a movement of the eyeball, the visual intake is limited before the closing and after the opening~\cite{volkmann1980eyeblinks}. Therefore, this part has to be marked in the eye tracking data~\cite{holmqvist2011eye}. Another interesting event measurable in modern high-speed eye trackers are post-saccadic oscillations. The cornea, a crystalline lens, is deformed during a saccade, thus influencing the pupil center estimation through the distortion of the pupil~\cite{hooge2015art,nystrom2015influence,tabernero2014lens}. This event is not grouped to fixations nor saccades \cite{nystrom2010adaptive}. During this event the subject can perceive but with distortions~\cite{tabernero2014lens}.

Up to now the choice of detection algorithm and parameters is up to the researcher, relying on literature values or the unfeasible task to annotate the data manually. The laborious and difficult task when using an algorithm is to adjust its parameters. Unfortunately, theoretically this process has to be repeated multiple times as the quality of the eye-tracking data often varies from subject to subject and between different tasks. If researchers want to analyze novel eye movements for which no gold standard algorithm exists, they have no choice but to annotate the data manually. With the proposed approach it is possible to create detectors theoretically even for yet unknown eye movements. Therefore, we propose to create detectors based on randomly generated binary decisions. We included ten different types of binary decision of which the final detector selects sets. Those sets learn a conditional distribution and can be combined to a single detector. This machine learning approach is random ferns~\cite{ozuysal2010fast}. We also propose an eye movement simulator to generate data similar to the data of the eye-tracker and to create a detector based on the simulation. This also enables to create detectors for very specific events such as skewed saccades.

\section{Related work}

\subsection{Eye movement simulation}
While there is well-funded knowledge about the gaze signal itself, its synthesis is still challenging. In the Eyecatch~\cite{yeo2012eyecatch} simulator, a Kalman filter is used to produce a gaze signal for saccades and smooth pursuits. While the signal itself was similar to real eye-tracking recordings, the jitter was missing. The first approach for rendering realistic and dynamic eye movements was proposed in \cite{lee2002eyes}, where the main focus was on saccadic eye movements. It also included smooth pursuits, binocular rotations (vergence) and the combination of eye and head rotations. The first data-driven approaches where proposed in \cite{ma2009natural} and \cite{peters2010head}. Both simulate head and eye movements together in order to generate eye-tracking data. The main disadvantage of \cite{ma2009natural} was that head motion seemed to trigger eye movement. In fact, the head orientation is only changed if the necessary amplitude of the eye is larger than a specific threshold~\cite{murphy2002perceptual} ($\approx 30^\circ$). Another data-driven approach was proposed in \cite{le2012live}, where an automated framework for head motion, gaze and eyelid simulation was developed. The framework generates data based on speech input using trained Gaussian Mixture Models. While this approach is capable of synthesizing non linear data, it only generates unperturbed gaze directions. The approach in \cite{duchowski2015modeling} models eye rotations using specific eye related quaternions for oculomotor rotations as proposed in \cite{tweed1990computing}. The main disadvantage of this approach is that the synthetic eyes cannot be rotated automatically. The approach in \cite{wood2015rendering} produces gaze vectors and eye images to train machine learning approaches for gaze prediction, but does not synthesize realistic eye movements.

All of the afore mentioned approaches have their origin in computer graphics with the goal to generate visually realistic head movement and gaze data. The main application of those simulators are to produce realistic interacting virtual humans using parametric models~\cite{andrist2012designing,pejsa2013stylized}. This leads to the disadvantage, that all movements in the generated data are perfect optimal representatives. In reality an important part of natural eye movements is noise introduces either through actual movements such as microsaccades or inaccuracies of the used eye-tracker. The first approach to simulate a realistic scanpath, i.e., a sequence of fixations and saccades, on static images was proposed in \cite{campbell2014saliency}. They use a saliency map together with a unified Bayesian model to generate realistic random walks over a stimulus. A pure gaze data simulation approach including noise was proposed in \cite{duchowski2015eye}. Based on this approach, \cite{duchowski2016eye} further improves the noise synthesis by simulating jitter as a normal distribution.

\subsection{Detection algorithms}
The most prominent fixation and saccade detection algorithm is Identification by Dispersion-Threshold (IDT)~\cite{salvucci2000identifying}. It uses the data reduction proposed in \cite{widdel1984operational}. The algorithm uses two thresholds, one is for the maximum fixation dispersion and the other for the minimum fixation duration. Another simple to implement algorithm is the Identification by Velocity Threshold (IVT) ~\cite{salvucci2000identifying}, where each sample below a chosen velocity threshold is classified as fixation and above as saccade. It is mostly applicable for high speed recordings. Based on the IVT algorithm, a self-adaptive approach was proposed in \cite{engbert2003microsaccades,engbert2006microsaccades}, where it was developed to detect microsaccades. In that approach, the velocity threshold is automatically adapted to the noise level in the eye-tracking data. An algorithm especially designed to cope with noisy data is the Identification by Kalman Filter (IKF) algorithm~\cite{komogortsev2009eye}. It uses the Kalman filter to predict the next sample value based on previous values. Therefore, it interpolates the data in an online fashion. For classification, two thresholds are used: one for the predicted value (velocity or distance) and one for the minimum fixation duration. Similar to this algorithm an implementation using the χ2-test instead of the Kalman filter was proposed in \cite{komogortsev2010standardization}. In \cite{veneri2011automatic}, the Covariance Dispersion Algorithm (CDT) was proposed. It is an improvement of the F-tests dispersion algorithm (FDT)~\cite{veneri2010eye}. The F-test measures if two data samples belong to the same class and due to the assumption that the data follows normal distributions it is sensitive to noise. The improvement by the covariance matrix is introduced to cope with this problem. The algorithm needs three thresholds, one for the variance, one for the co-variance, and a third threshold for the minimum duration. The identification by a Minimal Spanning Tree (IMST)~\cite{komogortsev2010standardization} creates a tree upon the data, where the samples represent the leafs. The goal is to select all samples with a minimum of branches given a connected graph (the data). Hidden Markov Models (HMM) have been proposed in \cite{komogortsev2010standardization,tafaj2012bayesian,kasneci2014applicability,santini2016bayesian} to separate fixations from saccades and even to detect smooth pursuits. The HMM consists of at least two states (fixation and saccade). For each new velocity sample the model decides whether it belongs to the current state (classification) or a state transition has occurred. After each sample, the model is updated to adapt to the data. The first algorithm that detects post saccadic movements was proposed in \cite{nystrom2010adaptive}. Based on the noise in the data the algorithm also adapts its velocity thresholds. The Binocular-Individual Threshold (BIT) algorithm~\cite{van2011defining} was also designed to detect small saccades in noisy data. Therefore, it applies its thresholds to the data of both eyes, following the ideas that both eyes have to perform the same movement. This algorithm also adapts its thresholds automatically. An algorithm detecting fixations, saccades, post saccadic movement and smooth pursuits was proposed in \cite{larsson2013detection}. This algorithms adapts the parameters automatically and is the first method capable of detecting all these eye movements at the same time. For high-speed eye-tracking data an algorithm for fixation, saccade and smooth pursuit detection was proposed in \cite{larsson2015detection}. The algorithm uses three stages to classify the data, starting with a preliminary segmentation and then evaluating each segment again, followed by the final classification. 

\section{Simulation}
 \begin{figure}[h]
	\centering\includegraphics[width=0.48\textwidth]{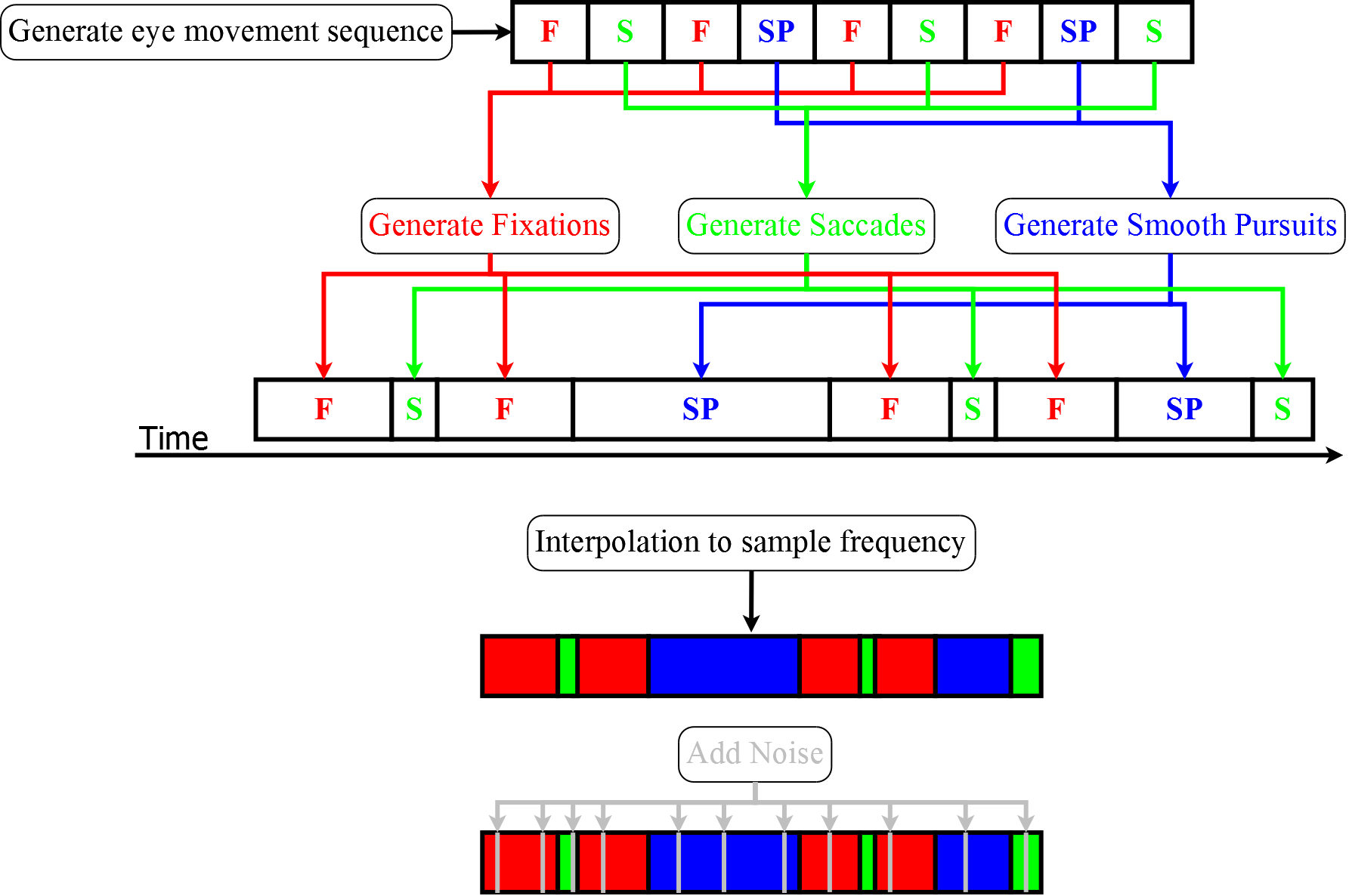}
	\caption{Work-flow of generating eye movement data. First, a sequence of eye movement types is generated. In the second step, a model of each eye movement type is generated (F:Fixation, S:Saccade, SM:Smooth pursuit). This model allows for an almost infinite sampling rate, which is in the next stage interpolated to a target sampling rate (Red:Fixation, Green:Saccade, Blue:Smooth pursuit). Finally, noise is added on top of the signal (gray). }
	\label{fig:workflow}
\end{figure}

The entire work-flow of the simulator is shown in Figure~\ref{fig:workflow}. Generating an eye movement velocity profile is done in four steps. The first step chooses a sequence of eye movement types (Fixation, Saccade, Smooth pursuit) without any time or velocity constrains. Afterwards, each movement type in this sequence is assigned a velocity profile generated by preliminary set parameters. The mathematical model behind these profiles allows sampling at an extremely high, almost arbitrary rate. The target sampling rate is obtained by interpolating this frequency, which also allows for dynamically adjusting the target sampling rate. In the last step, noise is added which represents measurement errors. Each step of this eye movement simulator is described in the following subsections in more detail.
The simulator also includes a random walker generator to model fixation direction~\cite{engbert2011integrated}; saccade and smooth pursuit directions are generated randomly (but consistently within a movement) since this are stimuli- and task-dependent.

\subsection{Eye movement sequence}
Generating a sequence of eye movement types can be done either by sampling from a uniform distribution, setting it manually, or by following construction constrains. In case of the uniform distributed eye movements, the generator script randomly selects between three types of eye movements. If the amount of each type is specified a priori, the probability is automatically adjusted. This means that after each insertion the probabilities are computed based on the remaining quantity of each type to favor higher quantities. This process can also be constrained, e.g., by forcing the algorithm to insert a saccade after each fixation or before a smooth pursuit.

\subsection{Fixation}
Fixations are generated based on two probability distributions which can be specified and parametrized. The first distribution determines its duration, the second the consistency of the fixation. For the duration and consistency the minimum and maximum can be set. As distributions, the simulator provides Normal and Uniform random number generation. For the Normal distribution, the standard deviation can be specified. consistency describes the fluctuations in the velocity profile and is used as such in the entire document.
 \begin{figure}[h]
	\centering
	\subfigure[]{\includegraphics[width=0.48\textwidth]{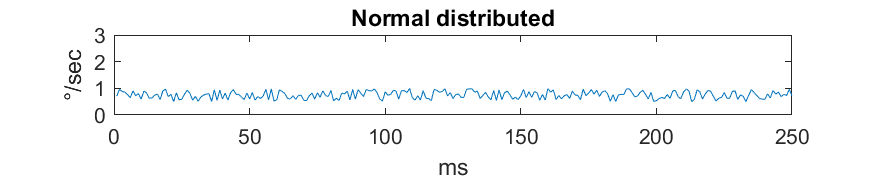}}

	\subfigure[]{\includegraphics[width=0.48\textwidth]{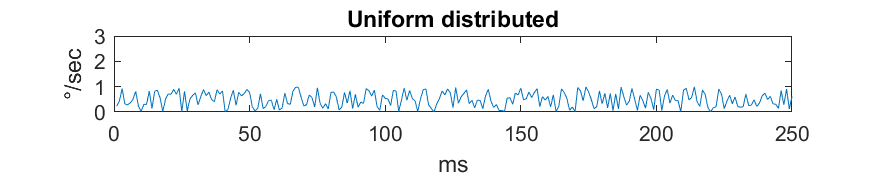}}
	\caption{Generated fixation based on a Normal (a) and Uniform (b) distribution.}
	\label{fig:fixexample}
\end{figure}

In Figure~\ref{fig:fixexample}, two artificially generated fixations are shown. The consistency was set to one degree per second and the standard deviation for the normal distribution to two
 (Figure~\ref{fig:fixexample} (a)). As can be seen in the figure, the Uniform distribution looks more similar compared to real data although we have set the consistency very high with one degree per second.

\subsection{Saccade}
The most complex part of the eye movement generator are the saccades. For the length, we follow the same approach as for the fixations, in which a minimum and maximum length has to be set. The selectable distributions are Normal and Uniform. The result of the length also influences the maximum speed of the saccade. Therefore, the two random numbers are multiplied (both in the range between zero and one).
This means that shorter saccades are limited to lower maximal velocities. To generate the velocity profile, minimum, maximum and the distribution tyoe have to be set.

The most characteristic property of a saccade is its velocity profile. In our simulator this is generated as a Gamma distribution. Therefore, the minimum and maximum skewness has to be specified. In \cite{van1987skewness} it was found that the Gamma function can be considered suitable to approximate saccade profiles (yet not perfect). To achieve more realistic data, a consistency minimum, maximum and distribution can be specified. This generates the jitter along the velocity profile.

 \begin{figure}[h]
	\centering
	\subfigure[]{\includegraphics[width=0.48\textwidth]{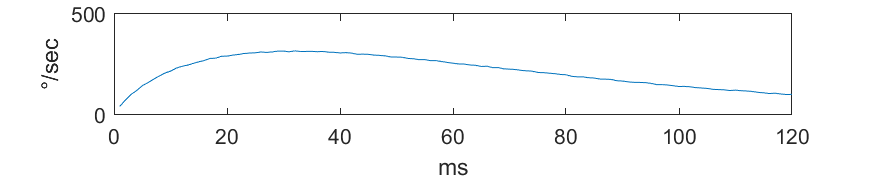}}

	\subfigure[]{\includegraphics[width=0.48\textwidth]{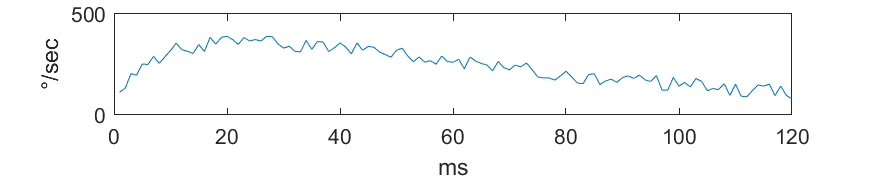}}

	\subfigure[]{\includegraphics[width=0.48\textwidth]{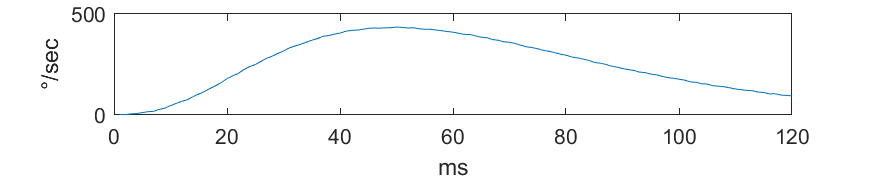}}

	\subfigure[]{\includegraphics[width=0.48\textwidth]{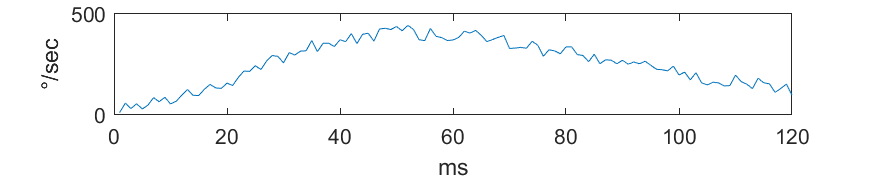}}
	\caption{Generated saccades with jitter (b,d) and without (a,c). For (a) and (b), the distribution was skewed to the left. In (c) and (d), Gamma distribution was only slightly skewed.}
	\label{fig:sacexample}
\end{figure}
Figure~\ref{fig:sacexample} shows some generated saccades of fixed length length. We simulated two large and two slightly left skewed saccades. The maximum velocity was selected from a range between $300$ and $500$ degrees per second.
As can be seen from the Figure, the profile contains on- and offset of a saccade. The profile itself is smooth and follows the Gamma distribution. Post-saccadic movement is as of now missing in the simulator. In Figure~\ref{fig:sacexample}(b) and (d), a small amount of jitter was added to simulate measurement inaccuracy. This usually occurs through the approximation on image pixels or ellipse fit inaccuracy in pupil detection.

\subsection{Smooth pursuit}
For generating smooth pursuits we also simulate the onset following the findings in \cite{ogawa1998velocity}. The authors did not provide a final function for the description of the velocity profile but visualized and described it precisely. The shape of the onset of a smooth pursuit follows a non linear growing function similar to the sigmoid function. While this equation is not scientifically proven, our framework allows to simply replace it once a better model is available. The most complex part of the pursuit model is the onset, followed by a regular movement.

The parameters that can be specified are the minimum and maximum length together with their distribution type. For the velocity and the length of the onset the same parameters can be adjusted. To include the measuring error, the consistency parameters are also configurable. For the pursuit itself we included linear growing, decreasing and constant profiles. In case of the growing, again the minimum, maximum and consistency function can be specified.
 \begin{figure}[h]
	\centering
	\subfigure[]{\includegraphics[width=0.22\textwidth]{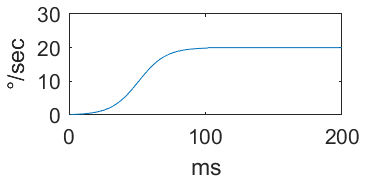}}
	\subfigure[]{\includegraphics[width=0.22\textwidth]{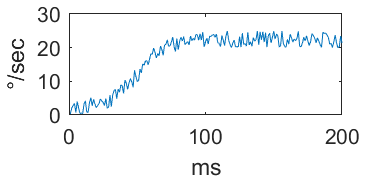}}

	\subfigure[]{\includegraphics[width=0.22\textwidth]{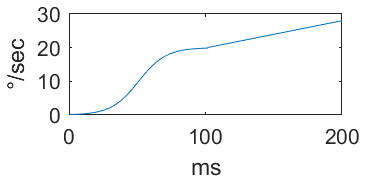}}
	\subfigure[]{\includegraphics[width=0.22\textwidth]{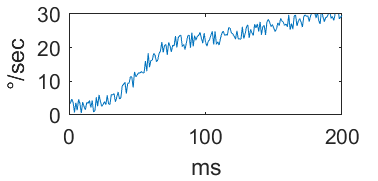}}

	\subfigure[]{\includegraphics[width=0.22\textwidth]{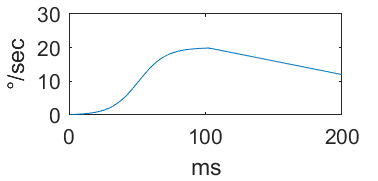}}
	\subfigure[]{\includegraphics[width=0.22\textwidth]{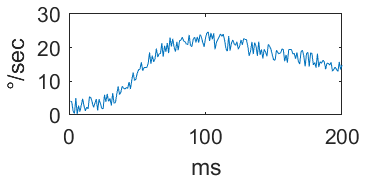}}
	\caption{Generated smooth pursuits with jitter (b,d,f) and without (a,c,e). For (a) and (b), the pursuit movement was constant. In (c,d) and (e,f) it was linear increasing and decreasing.}
	\label{fig:smexample}
\end{figure}
Figure~\ref{fig:smexample} shows simulated smooth pursuits. For the visualization of the linear decreasing and increasing function, extreme values were used. The first column shows a smooth pursuit for a constantly moving object, which is often observed in laboratory experiments. The increasing and decreasing profiles are for objects which move further away or come closer to the subject with a constant speed. Other profiles may occur in real settings too, where the object has a slightly varying speed but these are future extensions of the generator and not part of this paper.

\subsection{Sampling}
\label{sec:dwsampling}
After generating and linking the eye movements, they have to be interpolated to a sampling rate. This is necessary to simulate different recording frequencies. Here it is important to mention that not all modern eye trackers record at a constant frequency. On the one hand image acquisition rates can vary depending on illumination changes that affect the aperture time of the camera and timestamps generated by the eye-tracker can vary in accuracy. On the other hand, image processing time, e.g. for eye and pupil detection, are not necessarily constant and might change depending on how easy the pupil can be identified. For example, detection of the pupil is usually more time-consuming than keeping track of a previously detected pupil. Some systems, especially when running on mobile devices, may run into a state where frames are dropped in order to maintain real time performance. We found systems where the timestamps are generated by the CPU time (which may be inaccurate for fast sampling rates) and even timestamps that are generated after image processing. Therefore, our simulator is capable of simulating varying sampling rates. The parameters for this step are the minimum and maximum sampling rate and also the consistency function. The interpolation itself computes the mean of all values from the last sampling position to the new sampling position.

 \begin{figure}[h]
	\centering
	\subfigure[]{\includegraphics[width=0.48\textwidth]{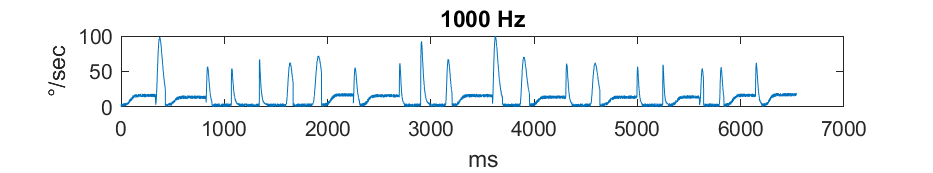}}

	\subfigure[]{\includegraphics[width=0.48\textwidth]{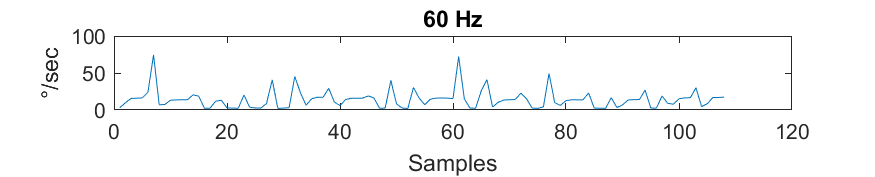}}

	\subfigure[]{\includegraphics[width=0.48\textwidth]{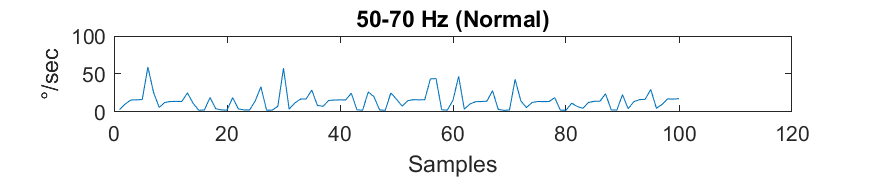}}

	\subfigure[]{\includegraphics[width=0.48\textwidth]{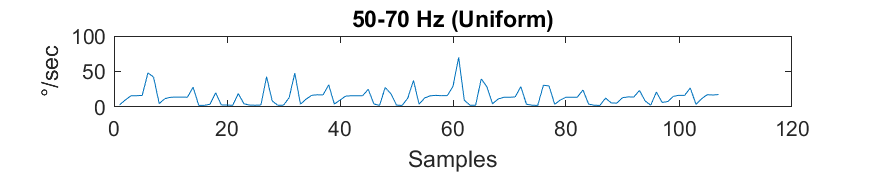}}
	\caption{Generated velocity profile of an eye movement sequence (a). In (b), the data is sampled at 60Hz without variations. (b) and (c) vary between 50 and 70 Hz with the Normal and the Uniform distribution.}
	\label{fig:pathexample}
\end{figure}
In Figure~\ref{fig:pathexample}(a) a generated velocity profile is shown. The initial sampling frequency was set to 1000 Hz but any other sampling rate is possible. For (b), a constant sampling frequency of 60 Hz was used. In (c), the sampling frequency varies between 50 and 70 Hz (with a mean of 60 Hz), wherein the Normal distribution was used as random number generator. It differs significantly from the constant sampling rate in (a) and also has a different length. For (d), the sampling frequency also varied between 50 and 70 Hz with the difference that the Uniform distribution was used as random number generator. The length is therefore similar to the constant sampling rate but it still differs especially for the saccadic peeks.

\subsection{Noise}
For generating noise, two distributions are used: one for the location where to place the noise in the data and the second for the velocity change to apply. Therefore, the user has to specify the types for both distributions and the minimum and maximum velocity of noise. The amount of noise is specified as a percentage of the samples that should be influenced.
 \begin{figure}[h]
	\centering
	\subfigure[]{\includegraphics[width=0.48\textwidth]{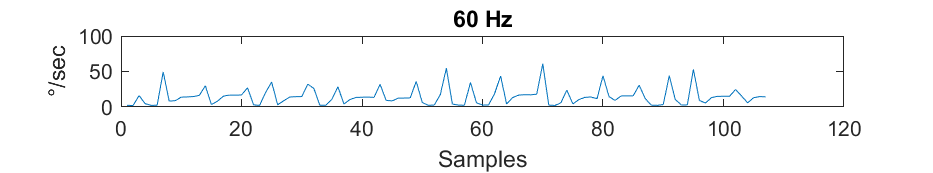}}

	\subfigure[]{\includegraphics[width=0.48\textwidth]{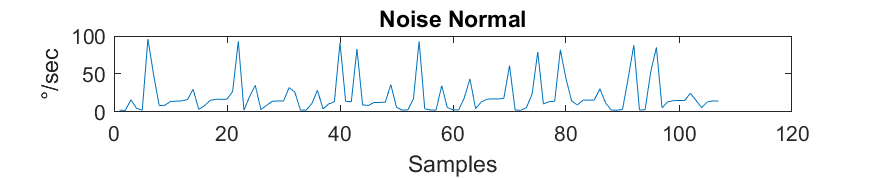}}

	\subfigure[]{\includegraphics[width=0.48\textwidth]{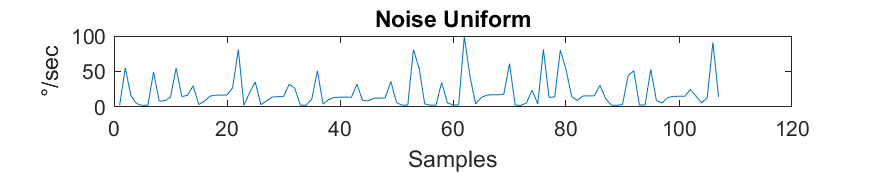}}
	\caption{Generated velocity profile of an eye movement sequence (a). In (b), noise is added based on a Normal distribution and in (c) a Uniform distribution was used.}
	\label{fig:noiseexample}
\end{figure}

Figure~\ref{fig:noiseexample} shows two types of Noise added to the velocity profile shown in (a). The amount of noise added was 10\%. For the Normal distributed noise in (b) it can be seen that the peaks are mostly high. In comparison to it, the Uniform distributed noise in (c) produces more peaks of different heights.

\section{Detector creation}
 In state-of-the-art algorithms there are thresholds for upper and lower limits as well as for ranges which have to be fulfilled. The main disadvantage is that those thresholds are difficult to adjust to new data, where the sampling rate is not constant or no time information is given~\cite{andersson2017one}. Another issue with those thresholds is that for some data they work very well while for more noisy data they do not work at all or need intensive preprocessing (such as smoothing filters and outliers detection). Our idea is to use the traditional thresholding approach but to adapt the algorithm to the data. The first step in our algorithm is to randomly generate different types of thresholding approaches and thresholds. The following binary decisions are generated:
 \begin{table}[h] \centering
 	\begin{tabular}{l r}
 	$|P1-P2|<TH1$ & $|P1-P2|>TH1$ \\
 	$P1<TH1\;and\;P2<TH2$ & $P1<TH1$ \\
 	$P1>TH1\;and\;P2>TH2$ & $P1>TH1$ \\
 	$P1<TH1\;and\;P2>TH2$ & $P2<TH2$\\
 	$P1>TH1\;and\;P2<TH2$ & $P2>TH2$,
	\end{tabular}
	\label{ite:binarylist}
\end{table}

\noindent
where $P1$ and $P2$ are two samples of the generated sequence, e.g., two velocities. These points are not required to be sequential, in fact $P1$ can be earlier or later than $P2$ in the sequence of samples. Their relative offsets to the sample that is currently in consideration for being classified is generated randomly. Therefore, the binary decisions consist of up to two distances to the current classification position and up to two thresholds ($TH1, TH2$). Based on the two distances, the sample positions $P1$ and $P2$ are computed. For those distances there is also the option to restrict them to samples preceding the current position, so that classification can be performed online, without the knowledge of future samples.
  \begin{figure}[h]
 	\centering
 	\includegraphics[width=0.3\textwidth]{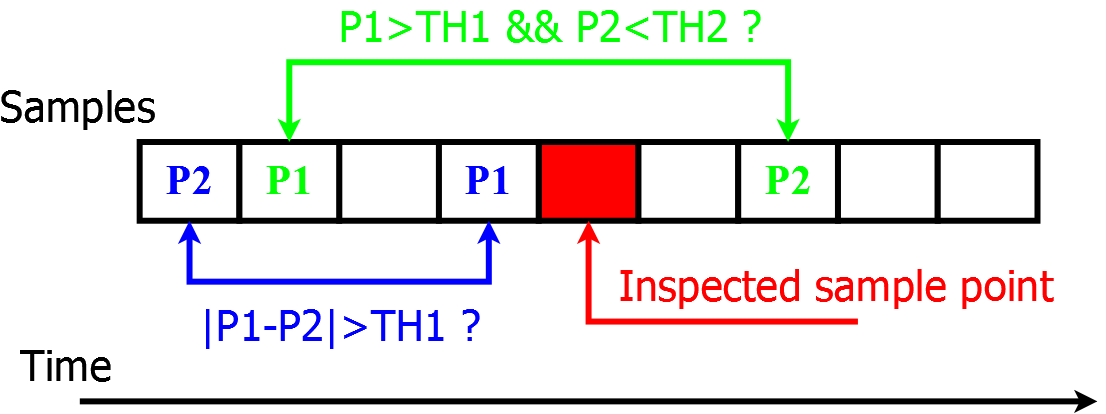}
 	\caption{Two binary decisions (green and blue) for the inspected sample position (red).}
 	\label{fig:binaryex}
 \end{figure}
 An example of such binary decisions is shown in Figure~\ref{fig:binaryex}. Red is the inspected sample for which a decision is to be made. The two binary decisions are colored blue and green. Blue is an example for an online applicable decision and green would be applied to an already complete recording or delayed to the recording. As can be seen, for each generated binary decision five parameters have to be selected (the fifth being the equation to use). A scan of all possible combinations of each parameter value would be far too extensive. Therefore, we generate hundreds of thousands randomly. Key to a good decision making is the selection of those parameter combinations that are relevant for determining an eye movement type. This means that we can assign a higher feature quality value to decisions that result in a true binary condition for a certain detection task, and vice versa. Afterwards, only the top ten percent with the highest quality values assigned to them are further used.

 The next step is to combine these binary decisions to a detector. This is done by computing a conditional distribution for a randomly selected set of ten binary decisions ($B$).
\begin{equation}
D(X,B) = p(X|B)
\label{eq:conprobdist}
\end{equation}
Equation~\ref{eq:conprobdist} describes this conditional distribution. $X$ is the current sample point, $B$ are the binary decisions and $p$ is the probability of the sample to be an eye movement type.
  \begin{figure}[h]
	\centering
	\includegraphics[width=0.3\textwidth]{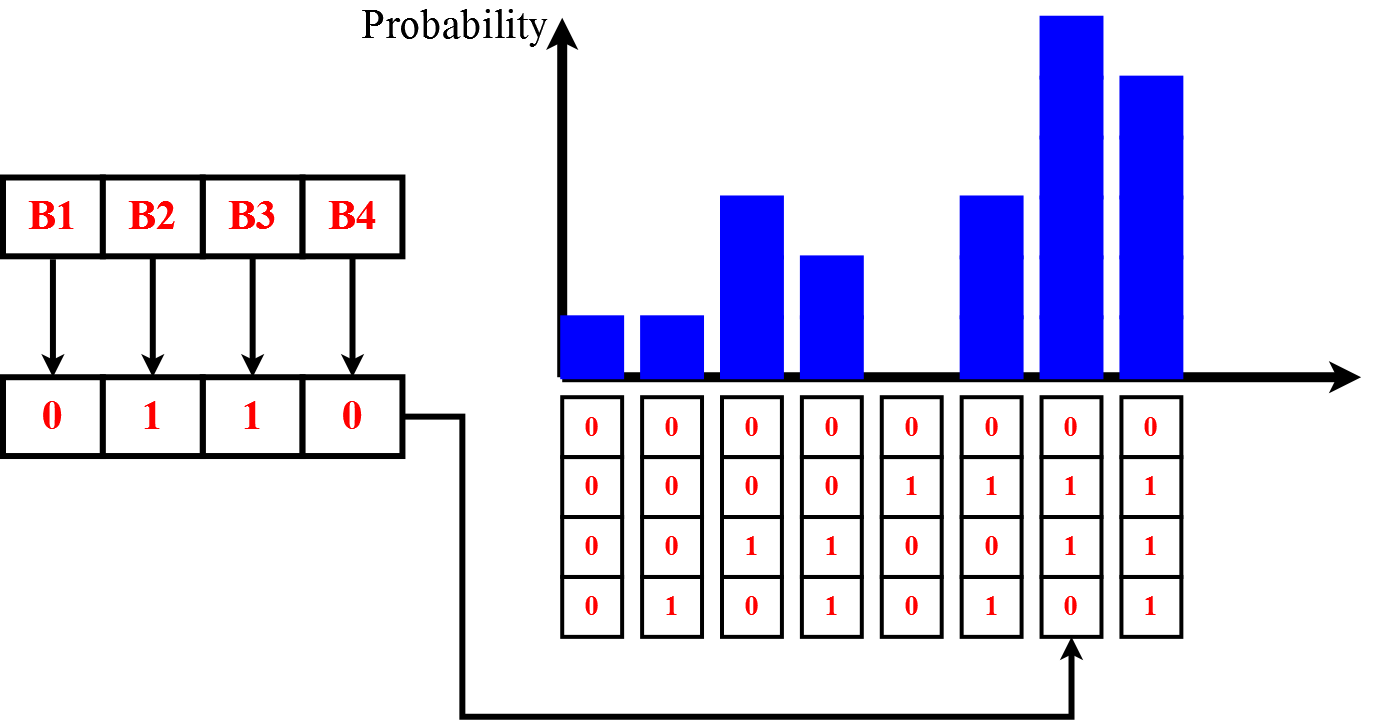}
	\caption{The binary decisions (red) represent an index in the distribution (blue) which holds the probability.}
	\label{fig:fernex}
\end{figure}
Figure~\ref{fig:fernex} shows the conditional distribution $p$ in blue. Each binary decision represents a digit, that is used as index in the conditional distribution. This combines multiple binary decisions to one detector. For training we use two distributions, one for valid examples and one for wrong examples. This allows to compute the final distribution without negative probabilities or to stop at zero. The difference between both distributions is the final score for a sample ($p_{positive}(X|B)-p_{negative}(X|B)$). The computation of each distribution is a simple lookup by the binary decisions index number and increasing the respective histogram index. The increment has to be normed to equalize the amount of positive and negative samples (the ratio between both occurrences in the training set).

After training of the conditional distributions for the so-called ferns, we have thousands of randomly selected weak detectors available. To create one strong detector we combine multiple of these, again randomly. Therefore, we compute a quality score for each fern similar as it was done for the binary decisions, and again consider only the top ten percent. Afterwards, we randomly select ten ferns and combine them under the independence assumption to one strong detector.
\begin{equation}
C(X,D_1,...,D_n) = D_1(X,B_1) * ... * D_n(X,B_n)
\label{eq:conprobdistclass}
\end{equation}
Equation~\ref{eq:conprobdistclass} describes this computation where $D_1,...,D_n$ are the ferns and $B_1,...,B_n$ are their binary decisions. For each eye movement type, we randomly generate hundreds of such detectors and score them as described for the binary decisions. After scoring, we select the top ten percent and evaluate them in combination. For combining the classifiers of different eye movement types, we consider only the combinations of equally ranked classifiers, e.g. the best fixation classifier with the best saccade classifier.
\begin{equation}
Type(X) =\begin{cases}
	Fixation & C_1 > C_2,C_3,C_4\\
	Saccade & C_2 > C_1,C_3,C_4 \\
	Smooth\; Pursuit & C_3 > C_1,C_2,C_4 \\
	Noise & C_4 > C_1,C_2,C_3
\end{cases}
\label{eq:detect}
\end{equation}
The final result is obtained using Equation~\ref{eq:detect}, where the highest probability decides which type is detected. The best combination of individual movement classifiers for the evaluation on the training set is selected as final detector. This also shows that it is easy to extend the detector for novel eye movement types or to train it only for specific events like the beginning of a movement or movement combinations such as regressions during reading.

\section{Evaluation}
For the entire evaluation we transformed the input signal to a sample per sample velocity signal. This was done to have an extremely challenging signal to simulate and for the detection. Most of the state-of-the-art algorithms apply smoothing or compute the velocity from multiple sequential samples. While such methods can always be applied the purpose of this evaluation is to show that our approach can adapt to the noise level even for challenging conditions. This sample per sample velocity signal is used in the entire evaluation. The evaluated data sets of annotated eye movements are chosen from \cite{santini2016bayesian,dorr2010variability,larsson2013detection} and contain multiple annotators. We evaluated each annotation separately, meaning that each annotator was evaluated as ground truth independent of the others.

The first subsection presents the evaluation of the proposed simulator, where examples are given showing real velocity profiles of recorded saccades from publicly available data sets. In the second part of this section, the trained detectors are evaluated and compared to state-of-the-art algorithms.

\subsection{Evaluation of the simulation}
 \begin{figure*}[ht]
	\centering
	\subfigure[]{\includegraphics[width=0.3\textwidth]{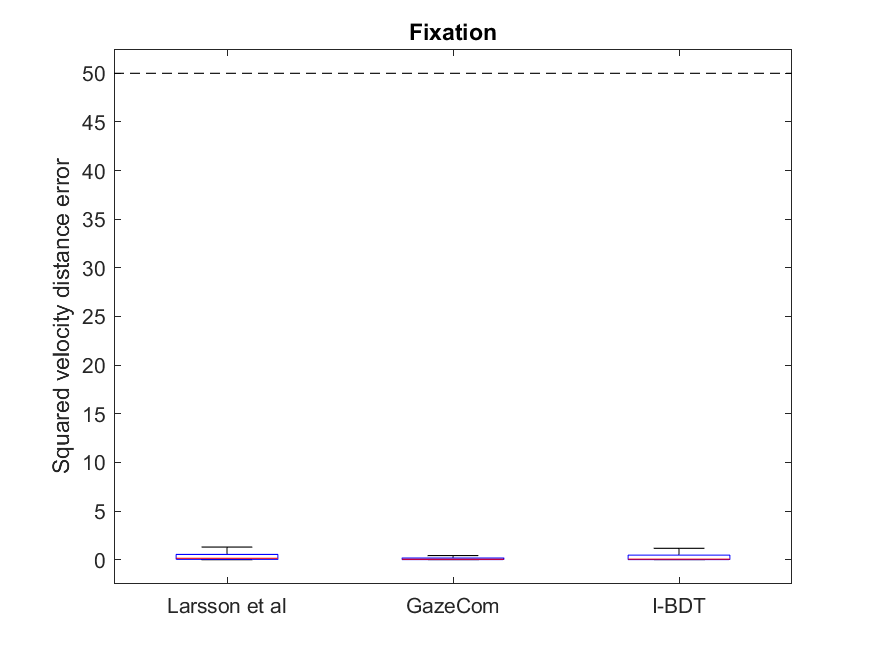}}
	\subfigure[]{\includegraphics[width=0.3\textwidth]{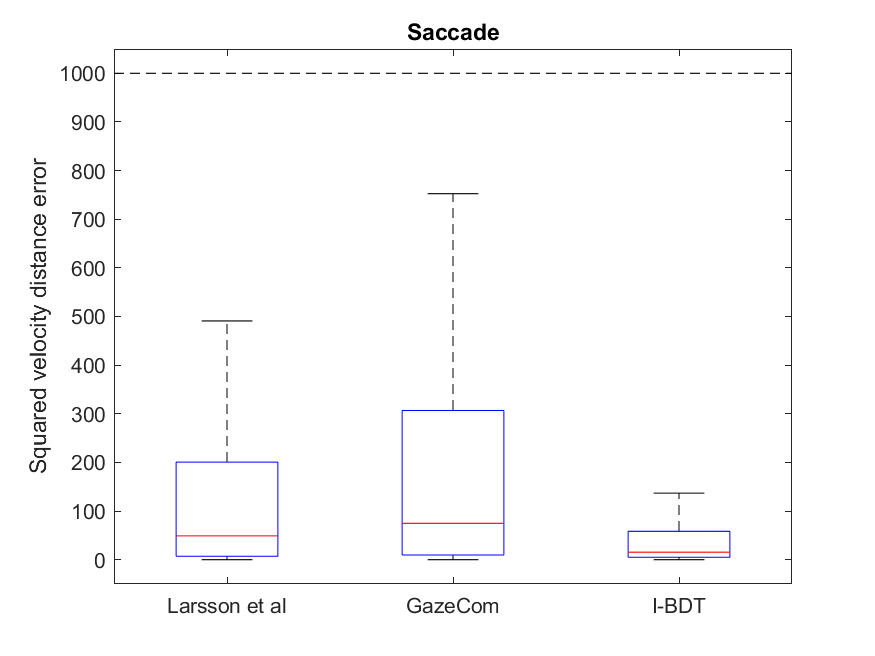}}
	\subfigure[]{\includegraphics[width=0.3\textwidth]{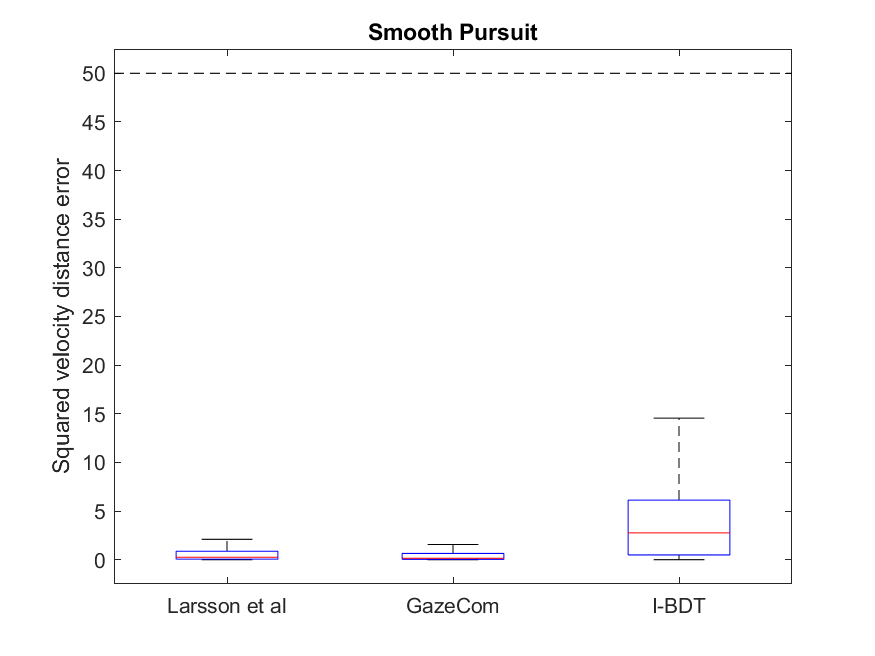}}
	\caption{Squared velocity error for the simulation per data set.}
	\label{fig:eval_simu}
\end{figure*}
Figure~\ref{fig:eval_simu} shows the per sample point squared error as whisker plots of our simulator in comparison to the publicly available data sets. The error was computed based on the squared difference between each sample. Therefore, we simulated each fixation, saccade, and smooth pursuit ten times with the same length as in the available data sets. For a fixation, the simulator got the information of the mean velocity and the standard deviation to generate a profile. The information of a saccade was the peak velocity and the position of this peak. For smooth pursuits, the simulator got the information of the mean velocity and the standard deviation. 

As can be seen in Figure~\ref{fig:eval_simu}(b), the error for saccades was the largest. This is due to the noisy signal which comes from the sample per sample velocity.
\begin{figure}[htb]
	\centering
	\subfigure[]{\includegraphics[width=0.15\textwidth]{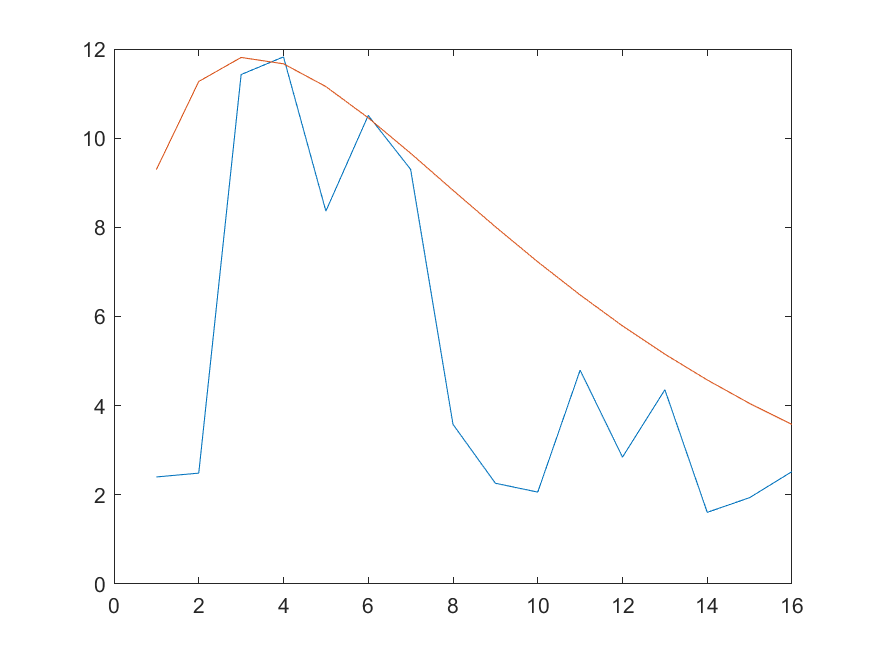}}
	\subfigure[]{\includegraphics[width=0.15\textwidth]{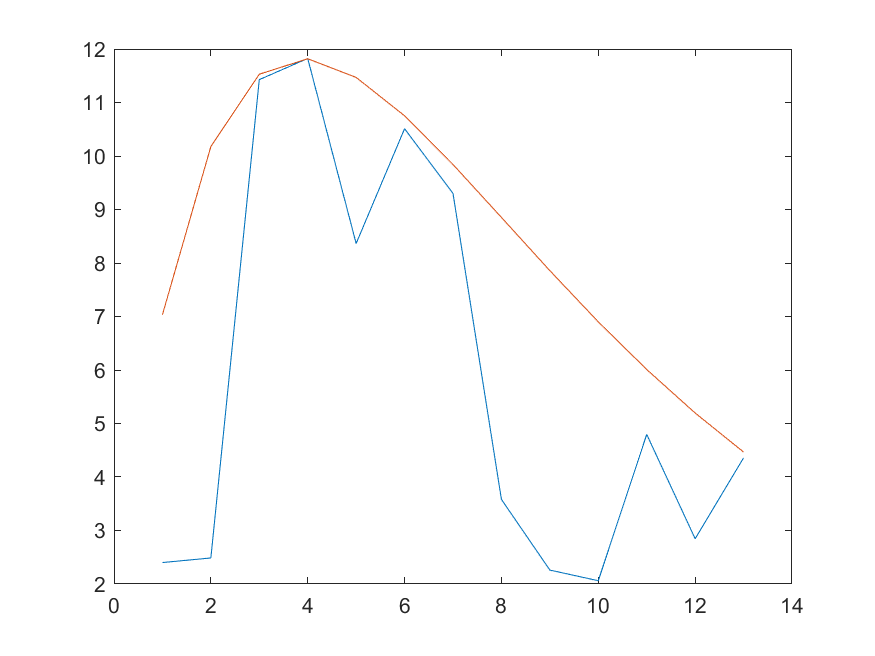}}
	\subfigure[]{\includegraphics[width=0.15\textwidth]{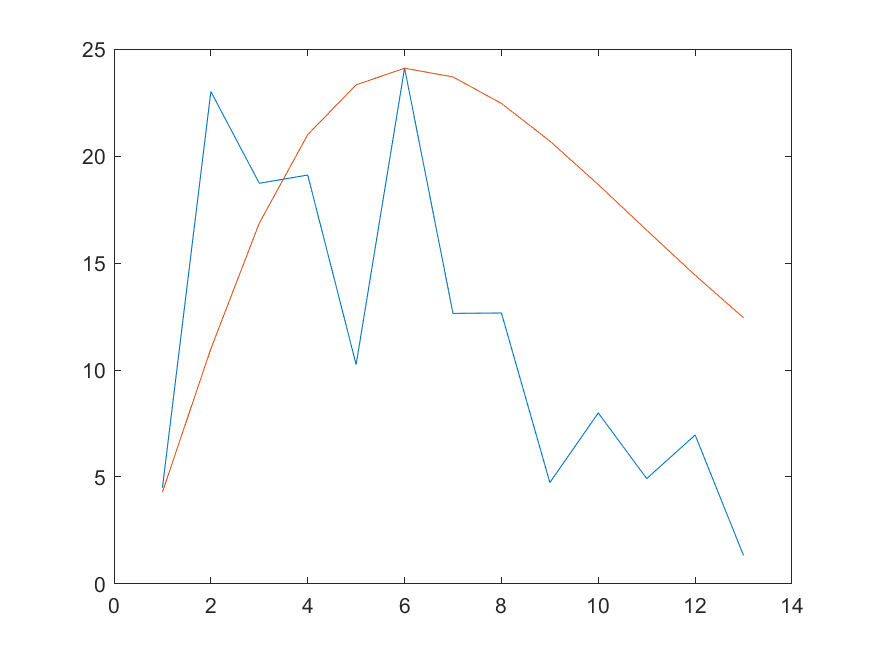}}
	\caption{Saccades with high squared error. Red is the simulation and blue the real data.}
	\label{fig:eval_sacc}
\end{figure}
Figure~\ref{fig:eval_sacc} shows some saccades which produced high squared errors. The red line corresponds to the simulation result, whereas the blue line corresponds to the real data. As can be seen, the course of the velocity profile is well simulated, which is well in line with previous findings in~\cite{van1987skewness}. The high errors originate mainly from measurement inaccuracies in the real data. This also highlights the difficulty in detecting eye movements in such a signal. For the data set from \cite{santini2016bayesian}(I-BDT), the error for saccades was lowest. This is due to the low sampling rate of the used eye tracker (30 Hz), for which large fluctuations do not occur. This is similar to smoothing or using multiple samples for the velocity computation. In contrast, the smooth pursuits error was the largest in the I-BDT data set. This is because in such low sampling rates the onset of a smooth pursuit is hardly represented. Our simulator is capable of simulating this (sampling~\ref{sec:dwsampling}) but for the evaluation it was not used. We only used the generators to simulate the eye movements.

\subsection{Evaluation of the detectors}
For comparison, we adapted the algorithms from \cite{santini2016bayesian} (I-BDT), \cite{nystrom2010adaptive} (EV), and \cite{larsson2013detection} (LS) to work with the velocity profile instead of x, y coordinates. We chose those algorithms because all come together with a self-adaptation procedure. I-BDT initializes itself on a part of the data and continues updating its probability distributions during runtime. For initialization, in the evaluation we provided the algorithm with the computed mean and standard deviation of the data it was evaluated on instead of the initial 15 seconds as done in the original implementation. The algorithm EV automatically adjusts its thresholds. Therefore, we provided it with the appropriate computed values for minimum velocity, maximum velocity, minimum duration, maximum duration, mean noise velocity etc. from the data it was evaluated on. LS is the representative for a segmentation-based self-adapting algorithm. We provided the statistical data similar to EV. As for I-BDT and EV, this means that for each evaluation we computed the statistics for the algorithms as initialization of their parameters based on the data they are evaluated on. This was done to simulate a handcrafted initialization. For the proposed approach we used simulated data to train and select a detector which was afterwards evaluated on the annotated data set. We did not use any post- or preprocessing of the data, nor segmentation or similar.

\begin{table}[h] \centering
	\begin{tabular}{llcccc}
		 \textbf{Data} & \textbf{Alg.} & \multicolumn{4}{c}{\textbf{Detection Rate (\%)}} \\
		 & & \textit{Fixation} & \textit{Saccade} & \textit{Pursuit} & \textit{Noise} \\ \hline
		\multirow{4}{*}{\rotatebox{90}{Larsson}} & EV & 92.47 & 4.37 & \phantom{0}0.00 & 35.11 \\
		& IBDT & \phantom{0}0.10 & \phantom{0}0.01 & 96.32 & \phantom{0}0.00 \\
		& LS & 23.73 & 72.69 & 56.54 & \phantom{0}3.98 \\
		& Proposed & 61.86 & 76.56 & 61.88 & 23.43 \\ \hline
		\multirow{4}{*}{\rotatebox{90}{GazeCom}} & EV & 94.86 & \phantom{0}0.78 & \phantom{0}0.00 & 43.19 \\
		& IBDT & \phantom{0}0.03 & \phantom{0}0.02 & 99.85 & \phantom{0}0.00 \\
		& LS & 49.18 & 70.79 & 43.18 & \phantom{0}5.86 \\
		& Proposed &  70.55 & 70.40 & 51.60 & 20.27 \\ \hline
		\multirow{4}{*}{\rotatebox{90}{I-BDT}} & EV & 92.47 & \phantom{0}0.61 & \phantom{0}0.00 & 71.46 \\
		& IBDT & 22.48 & 24.22 & 97.29 & \phantom{0}0.00 \\
		& LS & 29.50 & 13.04 & 93.14 & 10.44 \\
		& Proposed & 72.50 & 69.44 & 64.83 & 20.88
	\end{tabular}
\caption{Percentage of correctly detected samples per data set.}
\label{tbl:correct_classified}
\end{table}
Table~\ref{tbl:correct_classified} shows the correctly detected results per data set for all evaluated algorithms. As can be seen, our approach results in constantly balanced accuracies for fixations, saccades, and smooth pursuits for all data sets without having seen any of the real recordings. The other algorithms tend to prefer different types of eye movement. I-BDT for example cannot handle high speed recordings because the probability for smooth pursuits dominates. This is additionally supported by noise in the data which the algorithm is not designed to detect. The probably best performing competitor is LS~\cite{larsson2013detection}. For EV~\cite{nystrom2010adaptive} it has to be mentioned, that it does not detect smooth pursuits at all, significantly simplifying the classification problem as the probably hardest to classify class~\cite{komogortsev2013automated} (where velocities are \textit{somewhere in the middle}) is left out. Still noise and saccades are often confused.

The results of our detector could simply be improved by applying post processing to avoid too short or too long durations. The detectors could also be selected by evaluating combinations on the training set not only the ones which are ranked equally. Another improvement is to detect only the start and ending points of eye movements and set the data in between accordingly (similar to the segmentation in \cite{larsson2013detection}) but this is out of scope of this paper and will be part of further research.

\section{Conclusion}
We proposed a novel eye movement detection approach which is based on machine learning. It is capable of training detectors for specific eye movements and expendable to new findings. The detectors are capable of outperforming the state-of-the-art and adaptable to new challenges from new eye trackers. In addition the detectors can be trained for offline and online analysis enabling a second validation and refinement stage for eye movement detection. The underlying simulator, which generates the training data is based on scientific findings to generate the velocity profiles of eye movements. It is capable of simulating any static or dynamic sampling rate and allows to select different distributions for noise, sampling shift, eye tracker accuracy etc. Further research will be the extension of the simulator to be also capable of generating post saccadic, optokinetic and vestibulo-ocular movement. Additionally, a validation and correction extension will be developed to refine eye movement data based on known velocity profiles and validate gaze positions.

\bibliographystyle{ACM-Reference-Format}
\bibliography{sample-bibliography}


\begin{thebibliography}{49}


\ifx \showCODEN    \undefined \def \showCODEN     #1{\unskip}     \fi
\ifx \showDOI      \undefined \def \showDOI       #1{#1}\fi
\ifx \showISBNx    \undefined \def \showISBNx     #1{\unskip}     \fi
\ifx \showISBNxiii \undefined \def \showISBNxiii  #1{\unskip}     \fi
\ifx \showISSN     \undefined \def \showISSN      #1{\unskip}     \fi
\ifx \showLCCN     \undefined \def \showLCCN      #1{\unskip}     \fi
\ifx \shownote     \undefined \def \shownote      #1{#1}          \fi
\ifx \showarticletitle \undefined \def \showarticletitle #1{#1}   \fi
\ifx \showURL      \undefined \def \showURL       {\relax}        \fi
\providecommand\bibfield[2]{#2}
\providecommand\bibinfo[2]{#2}
\providecommand\natexlab[1]{#1}
\providecommand\showeprint[2][]{arXiv:#2}

\bibitem[\protect\citeauthoryear{Andersson, Larsson, Holmqvist, Stridh, and
  Nystr{\"o}m}{Andersson et~al\mbox{.}}{2017}]%
        {andersson2017one}
\bibfield{author}{\bibinfo{person}{Richard Andersson}, \bibinfo{person}{Linnea
  Larsson}, \bibinfo{person}{Kenneth Holmqvist}, \bibinfo{person}{Martin
  Stridh}, {and} \bibinfo{person}{Marcus Nystr{\"o}m}.}
  \bibinfo{year}{2017}\natexlab{}.
\newblock \showarticletitle{One algorithm to rule them all? An evaluation and
  discussion of ten eye movement event-detection algorithms}.
\newblock \bibinfo{journal}{{\em Behavior Research Methods\/}}
  \bibinfo{volume}{49}, \bibinfo{number}{2} (\bibinfo{year}{2017}),
  \bibinfo{pages}{616--637}.
\newblock


\bibitem[\protect\citeauthoryear{Andrist, Pejsa, Mutlu, and Gleicher}{Andrist
  et~al\mbox{.}}{2012}]%
        {andrist2012designing}
\bibfield{author}{\bibinfo{person}{Sean Andrist}, \bibinfo{person}{Tomislav
  Pejsa}, \bibinfo{person}{Bilge Mutlu}, {and} \bibinfo{person}{Michael
  Gleicher}.} \bibinfo{year}{2012}\natexlab{}.
\newblock \showarticletitle{Designing effective gaze mechanisms for virtual
  agents}. In \bibinfo{booktitle}{{\em Proceedings of the SIGCHI conference on
  Human Factors in Computing Systems}}. ACM, \bibinfo{pages}{705--714}.
\newblock


\bibitem[\protect\citeauthoryear{Braunagel, Geisler, Rosenstiel, and
  Kasneci}{Braunagel et~al\mbox{.}}{2017}]%
        {braunagel2017online}
\bibfield{author}{\bibinfo{person}{Christian Braunagel}, \bibinfo{person}{David
  Geisler}, \bibinfo{person}{Wolfgang Rosenstiel}, {and}
  \bibinfo{person}{Enkelejda Kasneci}.} \bibinfo{year}{2017}\natexlab{}.
\newblock \showarticletitle{Online Recognition of Driver-Activity Based on
  Visual Scanpath Classification}.
\newblock \bibinfo{journal}{{\em IEEE Intelligent Transportation Systems
  Magazine\/}} \bibinfo{volume}{9}, \bibinfo{number}{4} (\bibinfo{year}{2017}),
  \bibinfo{pages}{23--36}.
\newblock


\bibitem[\protect\citeauthoryear{Campbell, Chang, Chawarska, and Shic}{Campbell
  et~al\mbox{.}}{2014}]%
        {campbell2014saliency}
\bibfield{author}{\bibinfo{person}{Daniel~J Campbell}, \bibinfo{person}{Joseph
  Chang}, \bibinfo{person}{Katarzyna Chawarska}, {and}
  \bibinfo{person}{Frederick Shic}.} \bibinfo{year}{2014}\natexlab{}.
\newblock \showarticletitle{Saliency-based bayesian modeling of dynamic viewing
  of static scenes}. In \bibinfo{booktitle}{{\em Proceedings of the Symposium
  on Eye Tracking Research and Applications}}. ACM, \bibinfo{pages}{51--58}.
\newblock


\bibitem[\protect\citeauthoryear{Cornelissen, Peters, and Palmer}{Cornelissen
  et~al\mbox{.}}{2002}]%
        {cornelissen2002eyelink}
\bibfield{author}{\bibinfo{person}{Frans~W Cornelissen},
  \bibinfo{person}{Enno~M Peters}, {and} \bibinfo{person}{John Palmer}.}
  \bibinfo{year}{2002}\natexlab{}.
\newblock \showarticletitle{The Eyelink Toolbox: eye tracking with MATLAB and
  the Psychophysics Toolbox}.
\newblock \bibinfo{journal}{{\em Behavior Research Methods, Instruments, \&
  Computers\/}} \bibinfo{volume}{34}, \bibinfo{number}{4}
  (\bibinfo{year}{2002}), \bibinfo{pages}{613--617}.
\newblock


\bibitem[\protect\citeauthoryear{Dorr, Martinetz, Gegenfurtner, and Barth}{Dorr
  et~al\mbox{.}}{2010}]%
        {dorr2010variability}
\bibfield{author}{\bibinfo{person}{Michael Dorr}, \bibinfo{person}{Thomas
  Martinetz}, \bibinfo{person}{Karl~R Gegenfurtner}, {and}
  \bibinfo{person}{Erhardt Barth}.} \bibinfo{year}{2010}\natexlab{}.
\newblock \showarticletitle{Variability of eye movements when viewing dynamic
  natural scenes}.
\newblock \bibinfo{journal}{{\em Journal of vision\/}} \bibinfo{volume}{10},
  \bibinfo{number}{10} (\bibinfo{year}{2010}), \bibinfo{pages}{28--28}.
\newblock


\bibitem[\protect\citeauthoryear{Duchowski and J{\"o}rg}{Duchowski and
  J{\"o}rg}{2015}]%
        {duchowski2015modeling}
\bibfield{author}{\bibinfo{person}{AT Duchowski} {and} \bibinfo{person}{S
  J{\"o}rg}.} \bibinfo{year}{2015}\natexlab{}.
\newblock \showarticletitle{Modeling Physiologically Plausible Eye Rotations}.
\newblock \bibinfo{journal}{{\em Proceedings of Computer Graphics
  International\/}} (\bibinfo{year}{2015}).
\newblock


\bibitem[\protect\citeauthoryear{Duchowski, J{\"o}rg, Lawson, Bolte,
  {\'S}wirski, and Krejtz}{Duchowski et~al\mbox{.}}{2015}]%
        {duchowski2015eye}
\bibfield{author}{\bibinfo{person}{Andrew Duchowski}, \bibinfo{person}{Sophie
  J{\"o}rg}, \bibinfo{person}{Aubrey Lawson}, \bibinfo{person}{Takumi Bolte},
  \bibinfo{person}{Lech {\'S}wirski}, {and} \bibinfo{person}{Krzysztof
  Krejtz}.} \bibinfo{year}{2015}\natexlab{}.
\newblock \showarticletitle{Eye movement synthesis with 1/f pink noise}. In
  \bibinfo{booktitle}{{\em Proceedings of the 8th ACM SIGGRAPH Conference on
  Motion in Games}}. ACM, \bibinfo{pages}{47--56}.
\newblock


\bibitem[\protect\citeauthoryear{Duchowski}{Duchowski}{2002}]%
        {duchowski2002breadth}
\bibfield{author}{\bibinfo{person}{Andrew~T Duchowski}.}
  \bibinfo{year}{2002}\natexlab{}.
\newblock \showarticletitle{A breadth-first survey of eye-tracking
  applications}.
\newblock \bibinfo{journal}{{\em Behavior Research Methods, Instruments, \&
  Computers\/}} \bibinfo{volume}{34}, \bibinfo{number}{4}
  (\bibinfo{year}{2002}), \bibinfo{pages}{455--470}.
\newblock


\bibitem[\protect\citeauthoryear{Duchowski, J{\"o}rg, Allen, Giannopoulos, and
  Krejtz}{Duchowski et~al\mbox{.}}{2016}]%
        {duchowski2016eye}
\bibfield{author}{\bibinfo{person}{Andrew~T Duchowski}, \bibinfo{person}{Sophie
  J{\"o}rg}, \bibinfo{person}{Tyler~N Allen}, \bibinfo{person}{Ioannis
  Giannopoulos}, {and} \bibinfo{person}{Krzysztof Krejtz}.}
  \bibinfo{year}{2016}\natexlab{}.
\newblock \showarticletitle{Eye movement synthesis}. In
  \bibinfo{booktitle}{{\em Proceedings of the Symposium on Eye Tracking
  Research and Applications}}. ACM, \bibinfo{pages}{147--154}.
\newblock


\bibitem[\protect\citeauthoryear{Engbert and Kliegl}{Engbert and
  Kliegl}{2003}]%
        {engbert2003microsaccades}
\bibfield{author}{\bibinfo{person}{Ralf Engbert} {and}
  \bibinfo{person}{Reinhold Kliegl}.} \bibinfo{year}{2003}\natexlab{}.
\newblock \showarticletitle{Microsaccades uncover the orientation of covert
  attention}.
\newblock \bibinfo{journal}{{\em Vision research\/}} \bibinfo{volume}{43},
  \bibinfo{number}{9} (\bibinfo{year}{2003}), \bibinfo{pages}{1035--1045}.
\newblock


\bibitem[\protect\citeauthoryear{Engbert and Mergenthaler}{Engbert and
  Mergenthaler}{2006}]%
        {engbert2006microsaccades}
\bibfield{author}{\bibinfo{person}{Ralf Engbert} {and}
  \bibinfo{person}{Konstantin Mergenthaler}.} \bibinfo{year}{2006}\natexlab{}.
\newblock \showarticletitle{Microsaccades are triggered by low retinal image
  slip}.
\newblock \bibinfo{journal}{{\em Proceedings of the National Academy of
  Sciences\/}} \bibinfo{volume}{103}, \bibinfo{number}{18}
  (\bibinfo{year}{2006}), \bibinfo{pages}{7192--7197}.
\newblock


\bibitem[\protect\citeauthoryear{Engbert, Mergenthaler, Sinn, and
  Pikovsky}{Engbert et~al\mbox{.}}{2011}]%
        {engbert2011integrated}
\bibfield{author}{\bibinfo{person}{Ralf Engbert}, \bibinfo{person}{Konstantin
  Mergenthaler}, \bibinfo{person}{Petra Sinn}, {and} \bibinfo{person}{Arkady
  Pikovsky}.} \bibinfo{year}{2011}\natexlab{}.
\newblock \showarticletitle{An integrated model of fixational eye movements and
  microsaccades}.
\newblock \bibinfo{journal}{{\em Proceedings of the National Academy of
  Sciences\/}} \bibinfo{volume}{108}, \bibinfo{number}{39}
  (\bibinfo{year}{2011}), \bibinfo{pages}{E765--E770}.
\newblock


\bibitem[\protect\citeauthoryear{Holmqvist, Nystr{\"o}m, Andersson, Dewhurst,
  Jarodzka, and Van~de Weijer}{Holmqvist et~al\mbox{.}}{2011}]%
        {holmqvist2011eye}
\bibfield{author}{\bibinfo{person}{Kenneth Holmqvist}, \bibinfo{person}{Marcus
  Nystr{\"o}m}, \bibinfo{person}{Richard Andersson}, \bibinfo{person}{Richard
  Dewhurst}, \bibinfo{person}{Halszka Jarodzka}, {and} \bibinfo{person}{Joost
  Van~de Weijer}.} \bibinfo{year}{2011}\natexlab{}.
\newblock \bibinfo{booktitle}{{\em Eye tracking: A comprehensive guide to
  methods and measures}}.
\newblock \bibinfo{publisher}{OUP Oxford}.
\newblock


\bibitem[\protect\citeauthoryear{Hooge, Nystr{\"o}m, Cornelissen, and
  Holmqvist}{Hooge et~al\mbox{.}}{2015}]%
        {hooge2015art}
\bibfield{author}{\bibinfo{person}{Ignace Hooge}, \bibinfo{person}{Marcus
  Nystr{\"o}m}, \bibinfo{person}{Tim Cornelissen}, {and}
  \bibinfo{person}{Kenneth Holmqvist}.} \bibinfo{year}{2015}\natexlab{}.
\newblock \showarticletitle{The art of braking: Post saccadic oscillations in
  the eye tracker signal decrease with increasing saccade size}.
\newblock \bibinfo{journal}{{\em Vision research\/}}  \bibinfo{volume}{112}
  (\bibinfo{year}{2015}), \bibinfo{pages}{55--67}.
\newblock


\bibitem[\protect\citeauthoryear{Kasneci, Kasneci, K{\"u}bler, and
  Rosenstiel}{Kasneci et~al\mbox{.}}{2014}]%
        {kasneci2014applicability}
\bibfield{author}{\bibinfo{person}{Enkelejda Kasneci}, \bibinfo{person}{Gjergji
  Kasneci}, \bibinfo{person}{Thomas~C K{\"u}bler}, {and}
  \bibinfo{person}{Wolfgang Rosenstiel}.} \bibinfo{year}{2014}\natexlab{}.
\newblock \showarticletitle{The applicability of probabilistic methods to the
  online recognition of fixations and saccades in dynamic scenes}. In
  \bibinfo{booktitle}{{\em Proceedings of the Symposium on Eye Tracking
  Research and Applications}}. ACM, \bibinfo{pages}{323--326}.
\newblock


\bibitem[\protect\citeauthoryear{Kliegl and Olson}{Kliegl and Olson}{1981}]%
        {kliegl1981reduction}
\bibfield{author}{\bibinfo{person}{Reinhold Kliegl} {and}
  \bibinfo{person}{Richard~K Olson}.} \bibinfo{year}{1981}\natexlab{}.
\newblock \showarticletitle{Reduction and calibration of eye monitor data}.
\newblock \bibinfo{journal}{{\em Behavior Research Methods\/}}
  \bibinfo{volume}{13}, \bibinfo{number}{2} (\bibinfo{year}{1981}),
  \bibinfo{pages}{107--111}.
\newblock


\bibitem[\protect\citeauthoryear{Komogortsev, Gobert, Jayarathna, Koh, and
  Gowda}{Komogortsev et~al\mbox{.}}{2010}]%
        {komogortsev2010standardization}
\bibfield{author}{\bibinfo{person}{Oleg~V Komogortsev},
  \bibinfo{person}{Denise~V Gobert}, \bibinfo{person}{Sampath Jayarathna},
  \bibinfo{person}{Do~Hyong Koh}, {and} \bibinfo{person}{Sandeep~M Gowda}.}
  \bibinfo{year}{2010}\natexlab{}.
\newblock \showarticletitle{Standardization of automated analyses of oculomotor
  fixation and saccadic behaviors}.
\newblock \bibinfo{journal}{{\em IEEE Transactions on Biomedical
  Engineering\/}} \bibinfo{volume}{57}, \bibinfo{number}{11}
  (\bibinfo{year}{2010}), \bibinfo{pages}{2635--2645}.
\newblock


\bibitem[\protect\citeauthoryear{Komogortsev and Karpov}{Komogortsev and
  Karpov}{2013}]%
        {komogortsev2013automated}
\bibfield{author}{\bibinfo{person}{Oleg~V Komogortsev} {and}
  \bibinfo{person}{Alex Karpov}.} \bibinfo{year}{2013}\natexlab{}.
\newblock \showarticletitle{Automated classification and scoring of smooth
  pursuit eye movements in the presence of fixations and saccades}.
\newblock \bibinfo{journal}{{\em Behavior research methods\/}}
  \bibinfo{volume}{45}, \bibinfo{number}{1} (\bibinfo{year}{2013}),
  \bibinfo{pages}{203--215}.
\newblock


\bibitem[\protect\citeauthoryear{Komogortsev and Khan}{Komogortsev and
  Khan}{2009}]%
        {komogortsev2009eye}
\bibfield{author}{\bibinfo{person}{Oleg~V Komogortsev} {and}
  \bibinfo{person}{Javed~I Khan}.} \bibinfo{year}{2009}\natexlab{}.
\newblock \showarticletitle{Eye movement prediction by oculomotor plant Kalman
  filter with brainstem control}.
\newblock \bibinfo{journal}{{\em Journal of Control Theory and Applications\/}}
  \bibinfo{volume}{7}, \bibinfo{number}{1} (\bibinfo{year}{2009}),
  \bibinfo{pages}{14--22}.
\newblock


\bibitem[\protect\citeauthoryear{K{\"u}bler, Rothe, Schiefer, Rosenstiel, and
  Kasneci}{K{\"u}bler et~al\mbox{.}}{2017}]%
        {kubler2017subsmatch}
\bibfield{author}{\bibinfo{person}{Thomas~C K{\"u}bler},
  \bibinfo{person}{Colleen Rothe}, \bibinfo{person}{Ulrich Schiefer},
  \bibinfo{person}{Wolfgang Rosenstiel}, {and} \bibinfo{person}{Enkelejda
  Kasneci}.} \bibinfo{year}{2017}\natexlab{}.
\newblock \showarticletitle{SubsMatch 2.0: Scanpath comparison and
  classification based on subsequence frequencies}.
\newblock \bibinfo{journal}{{\em Behavior Research Methods\/}}
  \bibinfo{volume}{49}, \bibinfo{number}{3} (\bibinfo{year}{2017}),
  \bibinfo{pages}{1048--1064}.
\newblock


\bibitem[\protect\citeauthoryear{Larsson, Nystr{\"o}m, Andersson, and
  Stridh}{Larsson et~al\mbox{.}}{2015}]%
        {larsson2015detection}
\bibfield{author}{\bibinfo{person}{Linn{\'e}a Larsson}, \bibinfo{person}{Marcus
  Nystr{\"o}m}, \bibinfo{person}{Richard Andersson}, {and}
  \bibinfo{person}{Martin Stridh}.} \bibinfo{year}{2015}\natexlab{}.
\newblock \showarticletitle{Detection of fixations and smooth pursuit movements
  in high-speed eye-tracking data}.
\newblock \bibinfo{journal}{{\em Biomedical Signal Processing and Control\/}}
  \bibinfo{volume}{18} (\bibinfo{year}{2015}), \bibinfo{pages}{145--152}.
\newblock


\bibitem[\protect\citeauthoryear{Larsson, Nystr{\"o}m, and Stridh}{Larsson
  et~al\mbox{.}}{2013}]%
        {larsson2013detection}
\bibfield{author}{\bibinfo{person}{Linn{\'e}a Larsson}, \bibinfo{person}{Marcus
  Nystr{\"o}m}, {and} \bibinfo{person}{Martin Stridh}.}
  \bibinfo{year}{2013}\natexlab{}.
\newblock \showarticletitle{Detection of saccades and postsaccadic oscillations
  in the presence of smooth pursuit}.
\newblock \bibinfo{journal}{{\em IEEE Transactions on Biomedical
  Engineering\/}} \bibinfo{volume}{60}, \bibinfo{number}{9}
  (\bibinfo{year}{2013}), \bibinfo{pages}{2484--2493}.
\newblock


\bibitem[\protect\citeauthoryear{Le, Ma, and Deng}{Le et~al\mbox{.}}{2012}]%
        {le2012live}
\bibfield{author}{\bibinfo{person}{Binh~H Le}, \bibinfo{person}{Xiaohan Ma},
  {and} \bibinfo{person}{Zhigang Deng}.} \bibinfo{year}{2012}\natexlab{}.
\newblock \showarticletitle{Live speech driven head-and-eye motion generators}.
\newblock \bibinfo{journal}{{\em IEEE transactions on Visualization and
  Computer Graphics\/}} \bibinfo{volume}{18}, \bibinfo{number}{11}
  (\bibinfo{year}{2012}), \bibinfo{pages}{1902--1914}.
\newblock


\bibitem[\protect\citeauthoryear{Lee, Badler, and Badler}{Lee
  et~al\mbox{.}}{2002}]%
        {lee2002eyes}
\bibfield{author}{\bibinfo{person}{Sooha~Park Lee}, \bibinfo{person}{Jeremy~B
  Badler}, {and} \bibinfo{person}{Norman~I Badler}.}
  \bibinfo{year}{2002}\natexlab{}.
\newblock \showarticletitle{Eyes alive}. In \bibinfo{booktitle}{{\em ACM
  Transactions on Graphics (TOG)}}, Vol.~\bibinfo{volume}{21}. ACM,
  \bibinfo{pages}{637--644}.
\newblock


\bibitem[\protect\citeauthoryear{Leigh and Zee}{Leigh and Zee}{2015}]%
        {leigh2015neurology}
\bibfield{author}{\bibinfo{person}{R~John Leigh} {and} \bibinfo{person}{David~S
  Zee}.} \bibinfo{year}{2015}\natexlab{}.
\newblock \bibinfo{booktitle}{{\em The neurology of eye movements}}.
  Vol.~\bibinfo{volume}{90}.
\newblock \bibinfo{publisher}{Oxford University Press, USA}.
\newblock


\bibitem[\protect\citeauthoryear{Ma and Deng}{Ma and Deng}{2009}]%
        {ma2009natural}
\bibfield{author}{\bibinfo{person}{Xiaohan Ma} {and} \bibinfo{person}{Zhigang
  Deng}.} \bibinfo{year}{2009}\natexlab{}.
\newblock \showarticletitle{Natural eye motion synthesis by modeling gaze-head
  coupling}. In \bibinfo{booktitle}{{\em IEEE Virtual Reality Conference}}.
  IEEE, \bibinfo{pages}{143--150}.
\newblock


\bibitem[\protect\citeauthoryear{Murphy and Duchowski}{Murphy and
  Duchowski}{2002}]%
        {murphy2002perceptual}
\bibfield{author}{\bibinfo{person}{Hunter Murphy} {and}
  \bibinfo{person}{Andrew~T Duchowski}.} \bibinfo{year}{2002}\natexlab{}.
\newblock \showarticletitle{Perceptual gaze extent \& level of detail in VR:
  looking outside the box}. In \bibinfo{booktitle}{{\em ACM SIGGRAPH conference
  Abstracts and Applications}}. ACM, \bibinfo{pages}{228--228}.
\newblock


\bibitem[\protect\citeauthoryear{Nystr{\"o}m, Andersson, Magnusson, Pansell,
  and Hooge}{Nystr{\"o}m et~al\mbox{.}}{2015}]%
        {nystrom2015influence}
\bibfield{author}{\bibinfo{person}{Marcus Nystr{\"o}m},
  \bibinfo{person}{Richard Andersson}, \bibinfo{person}{M{\aa}ns Magnusson},
  \bibinfo{person}{Tony Pansell}, {and} \bibinfo{person}{Ignace Hooge}.}
  \bibinfo{year}{2015}\natexlab{}.
\newblock \showarticletitle{The influence of crystalline lens accommodation on
  post-saccadic oscillations in pupil-based eye trackers}.
\newblock \bibinfo{journal}{{\em Vision research\/}}  \bibinfo{volume}{107}
  (\bibinfo{year}{2015}), \bibinfo{pages}{1--14}.
\newblock


\bibitem[\protect\citeauthoryear{Nystr{\"o}m and Holmqvist}{Nystr{\"o}m and
  Holmqvist}{2010}]%
        {nystrom2010adaptive}
\bibfield{author}{\bibinfo{person}{Marcus Nystr{\"o}m} {and}
  \bibinfo{person}{Kenneth Holmqvist}.} \bibinfo{year}{2010}\natexlab{}.
\newblock \showarticletitle{An adaptive algorithm for fixation, saccade, and
  glissade detection in eyetracking data}.
\newblock \bibinfo{journal}{{\em Behavior research methods\/}}
  \bibinfo{volume}{42}, \bibinfo{number}{1} (\bibinfo{year}{2010}),
  \bibinfo{pages}{188--204}.
\newblock


\bibitem[\protect\citeauthoryear{Ogawa and Fujita}{Ogawa and Fujita}{1998}]%
        {ogawa1998velocity}
\bibfield{author}{\bibinfo{person}{Tadashi Ogawa} {and}
  \bibinfo{person}{Masahiko Fujita}.} \bibinfo{year}{1998}\natexlab{}.
\newblock \showarticletitle{Velocity profile of smooth pursuit eye movements in
  humans: pursuit velocity increase linked with the initial saccade
  occurrence}.
\newblock \bibinfo{journal}{{\em Neuroscience research\/}}
  \bibinfo{volume}{31}, \bibinfo{number}{3} (\bibinfo{year}{1998}),
  \bibinfo{pages}{201--209}.
\newblock


\bibitem[\protect\citeauthoryear{Ozuysal, Calonder, Lepetit, and Fua}{Ozuysal
  et~al\mbox{.}}{2010}]%
        {ozuysal2010fast}
\bibfield{author}{\bibinfo{person}{Mustafa Ozuysal}, \bibinfo{person}{Michael
  Calonder}, \bibinfo{person}{Vincent Lepetit}, {and} \bibinfo{person}{Pascal
  Fua}.} \bibinfo{year}{2010}\natexlab{}.
\newblock \showarticletitle{Fast keypoint recognition using random ferns}.
\newblock \bibinfo{journal}{{\em IEEE transactions on Pattern Analysis and
  Machine Intelligence\/}} \bibinfo{volume}{32}, \bibinfo{number}{3}
  (\bibinfo{year}{2010}), \bibinfo{pages}{448--461}.
\newblock


\bibitem[\protect\citeauthoryear{Pejsa, Mutlu, and Gleicher}{Pejsa
  et~al\mbox{.}}{2013}]%
        {pejsa2013stylized}
\bibfield{author}{\bibinfo{person}{Tomislav Pejsa}, \bibinfo{person}{Bilge
  Mutlu}, {and} \bibinfo{person}{Michael Gleicher}.}
  \bibinfo{year}{2013}\natexlab{}.
\newblock \showarticletitle{Stylized and performative gaze for character
  animation}. In \bibinfo{booktitle}{{\em Computer Graphics Forum}},
  Vol.~\bibinfo{volume}{32}. Wiley Online Library, \bibinfo{pages}{143--152}.
\newblock


\bibitem[\protect\citeauthoryear{Peters and Qureshi}{Peters and
  Qureshi}{2010}]%
        {peters2010head}
\bibfield{author}{\bibinfo{person}{Christopher Peters} {and}
  \bibinfo{person}{Adam Qureshi}.} \bibinfo{year}{2010}\natexlab{}.
\newblock \showarticletitle{A head movement propensity model for animating gaze
  shifts and blinks of virtual characters}.
\newblock \bibinfo{journal}{{\em Computers and Graphics\/}}
  \bibinfo{volume}{34}, \bibinfo{number}{6} (\bibinfo{year}{2010}),
  \bibinfo{pages}{677--687}.
\newblock


\bibitem[\protect\citeauthoryear{Rashbass}{Rashbass}{1961}]%
        {rashbass1961relationship}
\bibfield{author}{\bibinfo{person}{C1 Rashbass}.}
  \bibinfo{year}{1961}\natexlab{}.
\newblock \showarticletitle{The relationship between saccadic and smooth
  tracking eye movements}.
\newblock \bibinfo{journal}{{\em The Journal of Physiology\/}}
  \bibinfo{volume}{159}, \bibinfo{number}{2} (\bibinfo{year}{1961}),
  \bibinfo{pages}{326--338}.
\newblock


\bibitem[\protect\citeauthoryear{Rayner}{Rayner}{1998}]%
        {rayner1998eye}
\bibfield{author}{\bibinfo{person}{Keith Rayner}.}
  \bibinfo{year}{1998}\natexlab{}.
\newblock \showarticletitle{Eye movements in reading and information
  processing: 20 years of research.}
\newblock \bibinfo{journal}{{\em Psychological bulletin\/}}
  \bibinfo{volume}{124}, \bibinfo{number}{3} (\bibinfo{year}{1998}),
  \bibinfo{pages}{372}.
\newblock


\bibitem[\protect\citeauthoryear{Salvucci and Goldberg}{Salvucci and
  Goldberg}{2000}]%
        {salvucci2000identifying}
\bibfield{author}{\bibinfo{person}{Dario~D Salvucci} {and}
  \bibinfo{person}{Joseph~H Goldberg}.} \bibinfo{year}{2000}\natexlab{}.
\newblock \showarticletitle{Identifying fixations and saccades in eye-tracking
  protocols}. In \bibinfo{booktitle}{{\em Proceedings of the Symposium on Eye
  Tracking Research and Applications}}. ACM, \bibinfo{pages}{71--78}.
\newblock


\bibitem[\protect\citeauthoryear{Santini, Fuhl, K{\"u}bler, and
  Kasneci}{Santini et~al\mbox{.}}{2016}]%
        {santini2016bayesian}
\bibfield{author}{\bibinfo{person}{Thiago Santini}, \bibinfo{person}{Wolfgang
  Fuhl}, \bibinfo{person}{Thomas K{\"u}bler}, {and} \bibinfo{person}{Enkelejda
  Kasneci}.} \bibinfo{year}{2016}\natexlab{}.
\newblock \showarticletitle{Bayesian identification of fixations, saccades, and
  smooth pursuits}. In \bibinfo{booktitle}{{\em Proceedings of the Symposium on
  Eye Tracking Research and Applications}}. ACM, \bibinfo{pages}{163--170}.
\newblock


\bibitem[\protect\citeauthoryear{Tabernero and Artal}{Tabernero and
  Artal}{2014}]%
        {tabernero2014lens}
\bibfield{author}{\bibinfo{person}{Juan Tabernero} {and} \bibinfo{person}{Pablo
  Artal}.} \bibinfo{year}{2014}\natexlab{}.
\newblock \showarticletitle{Lens oscillations in the human eye. Implications
  for post-saccadic suppression of vision}.
\newblock \bibinfo{journal}{{\em PloS one\/}} \bibinfo{volume}{9},
  \bibinfo{number}{4} (\bibinfo{year}{2014}), \bibinfo{pages}{e95764}.
\newblock


\bibitem[\protect\citeauthoryear{Tafaj, Kasneci, Rosenstiel, and Bogdan}{Tafaj
  et~al\mbox{.}}{2012}]%
        {tafaj2012bayesian}
\bibfield{author}{\bibinfo{person}{Enkelejda Tafaj}, \bibinfo{person}{Gjergji
  Kasneci}, \bibinfo{person}{Wolfgang Rosenstiel}, {and}
  \bibinfo{person}{Martin Bogdan}.} \bibinfo{year}{2012}\natexlab{}.
\newblock \showarticletitle{Bayesian online clustering of eye movement data}.
  In \bibinfo{booktitle}{{\em Proceedings of the Symposium on Eye Tracking
  Research and Applications}}. ACM, \bibinfo{pages}{285--288}.
\newblock


\bibitem[\protect\citeauthoryear{Tweed, Cadera, and Vilis}{Tweed
  et~al\mbox{.}}{1990}]%
        {tweed1990computing}
\bibfield{author}{\bibinfo{person}{Douglas Tweed}, \bibinfo{person}{Werner
  Cadera}, {and} \bibinfo{person}{Tutis Vilis}.}
  \bibinfo{year}{1990}\natexlab{}.
\newblock \showarticletitle{Computing three-dimensional eye position
  quaternions and eye velocity from search coil signals}.
\newblock \bibinfo{journal}{{\em Vision research\/}} \bibinfo{volume}{30},
  \bibinfo{number}{1} (\bibinfo{year}{1990}), \bibinfo{pages}{97--110}.
\newblock


\bibitem[\protect\citeauthoryear{van~der Lans, Wedel, and Pieters}{van~der Lans
  et~al\mbox{.}}{2011}]%
        {van2011defining}
\bibfield{author}{\bibinfo{person}{Ralf van~der Lans}, \bibinfo{person}{Michel
  Wedel}, {and} \bibinfo{person}{Rik Pieters}.}
  \bibinfo{year}{2011}\natexlab{}.
\newblock \showarticletitle{Defining eye-fixation sequences across individuals
  and tasks: the Binocular-Individual Threshold (BIT) algorithm}.
\newblock \bibinfo{journal}{{\em Behavior Research Methods\/}}
  \bibinfo{volume}{43}, \bibinfo{number}{1} (\bibinfo{year}{2011}),
  \bibinfo{pages}{239--257}.
\newblock


\bibitem[\protect\citeauthoryear{Van~Opstal and Van~Gisbergen}{Van~Opstal and
  Van~Gisbergen}{1987}]%
        {van1987skewness}
\bibfield{author}{\bibinfo{person}{AJ Van~Opstal} {and} \bibinfo{person}{JAM
  Van~Gisbergen}.} \bibinfo{year}{1987}\natexlab{}.
\newblock \showarticletitle{Skewness of saccadic velocity profiles: a unifying
  parameter for normal and slow saccades}.
\newblock \bibinfo{journal}{{\em Vision research\/}} \bibinfo{volume}{27},
  \bibinfo{number}{5} (\bibinfo{year}{1987}), \bibinfo{pages}{731--745}.
\newblock


\bibitem[\protect\citeauthoryear{Veneri, Piu, Federighi, Rosini, Federico, and
  Rufa}{Veneri et~al\mbox{.}}{2010}]%
        {veneri2010eye}
\bibfield{author}{\bibinfo{person}{Giacomo Veneri}, \bibinfo{person}{Pietro
  Piu}, \bibinfo{person}{Pamela Federighi}, \bibinfo{person}{Francesca Rosini},
  \bibinfo{person}{Antonio Federico}, {and} \bibinfo{person}{Alessandra Rufa}.}
  \bibinfo{year}{2010}\natexlab{}.
\newblock \showarticletitle{Eye fixations identification based on statistical
  analysis-case study}. In \bibinfo{booktitle}{{\em Cognitive Information
  Processing (CIP), 2010 2nd International Workshop on}}. IEEE,
  \bibinfo{pages}{446--451}.
\newblock


\bibitem[\protect\citeauthoryear{Veneri, Piu, Rosini, Federighi, Federico, and
  Rufa}{Veneri et~al\mbox{.}}{2011}]%
        {veneri2011automatic}
\bibfield{author}{\bibinfo{person}{Giacomo Veneri}, \bibinfo{person}{Pietro
  Piu}, \bibinfo{person}{Francesca Rosini}, \bibinfo{person}{Pamela Federighi},
  \bibinfo{person}{Antonio Federico}, {and} \bibinfo{person}{Alessandra Rufa}.}
  \bibinfo{year}{2011}\natexlab{}.
\newblock \showarticletitle{Automatic eye fixations identification based on
  analysis of variance and covariance}.
\newblock \bibinfo{journal}{{\em Pattern Recognition Letters\/}}
  \bibinfo{volume}{32}, \bibinfo{number}{13} (\bibinfo{year}{2011}),
  \bibinfo{pages}{1588--1593}.
\newblock


\bibitem[\protect\citeauthoryear{Volkmann, Riggs, and Moore}{Volkmann
  et~al\mbox{.}}{1980}]%
        {volkmann1980eyeblinks}
\bibfield{author}{\bibinfo{person}{Frances~C Volkmann},
  \bibinfo{person}{Lorrin~A Riggs}, {and} \bibinfo{person}{Robert~K Moore}.}
  \bibinfo{year}{1980}\natexlab{}.
\newblock \showarticletitle{Eyeblinks and visual suppression}.
\newblock \bibinfo{journal}{{\em Science\/}} \bibinfo{volume}{207},
  \bibinfo{number}{4433} (\bibinfo{year}{1980}), \bibinfo{pages}{900--902}.
\newblock


\bibitem[\protect\citeauthoryear{Widdel}{Widdel}{1984}]%
        {widdel1984operational}
\bibfield{author}{\bibinfo{person}{Heino Widdel}.}
  \bibinfo{year}{1984}\natexlab{}.
\newblock \showarticletitle{Operational problems in analysing eye movements}.
\newblock \bibinfo{journal}{{\em Advances in psychology\/}}
  \bibinfo{volume}{22} (\bibinfo{year}{1984}), \bibinfo{pages}{21--29}.
\newblock


\bibitem[\protect\citeauthoryear{Wood, Baltrusaitis, Zhang, Sugano, Robinson,
  and Bulling}{Wood et~al\mbox{.}}{2015}]%
        {wood2015rendering}
\bibfield{author}{\bibinfo{person}{Erroll Wood}, \bibinfo{person}{Tadas
  Baltrusaitis}, \bibinfo{person}{Xucong Zhang}, \bibinfo{person}{Yusuke
  Sugano}, \bibinfo{person}{Peter Robinson}, {and} \bibinfo{person}{Andreas
  Bulling}.} \bibinfo{year}{2015}\natexlab{}.
\newblock \showarticletitle{Rendering of eyes for eye-shape registration and
  gaze estimation}. In \bibinfo{booktitle}{{\em Proceedings of the IEEE
  International Conference on Computer Vision}}. \bibinfo{pages}{3756--3764}.
\newblock


\bibitem[\protect\citeauthoryear{Yeo, Lesmana, Neog, and Pai}{Yeo
  et~al\mbox{.}}{2012}]%
        {yeo2012eyecatch}
\bibfield{author}{\bibinfo{person}{Sang~Hoon Yeo}, \bibinfo{person}{Martin
  Lesmana}, \bibinfo{person}{Debanga~R Neog}, {and} \bibinfo{person}{Dinesh~K
  Pai}.} \bibinfo{year}{2012}\natexlab{}.
\newblock \showarticletitle{Eyecatch: simulating visuomotor coordination for
  object interception}.
\newblock \bibinfo{journal}{{\em ACM Transactions on Graphics (TOG)\/}}
  \bibinfo{volume}{31}, \bibinfo{number}{4} (\bibinfo{year}{2012}),
  \bibinfo{pages}{42}.
\newblock


\end{thebibliography}

\end{document}